\documentclass[preprintnumbers, nofootinbib, preprint, 12pt]{revtex4}
\usepackage{amsmath,amssymb,bm}
\usepackage[dvipdfm]{graphicx}
\begin{document}
{~}
\vspace{3cm}

\title{Dynamical Black Rings with a Positive Cosmological Constant
\vspace{1cm}
}
\author{
Masashi Kimura\footnote{E-mail:mkimura@sci.osaka-cu.ac.jp}
}
\affiliation{
Department of Mathematics and Physics,  
Graduate School of Science, Osaka City University,
3-3-138 Sugimoto, Sumiyoshi, Osaka 558-8585, Japan
\vspace{3cm}
}

\begin{abstract}
We construct dynamical black ring solutions
in the five dimensional Einstein-Maxwell system with a positive cosmological constant
and investigate the geometrical structure.
The solutions describe the physical process such that
a thin black ring at early time 
shrinks  and changes into a single black hole as time increase.
We also discuss the multi-black rings and the coalescence of them.
\end{abstract}

\preprint{OCU-PHYS 312}
\preprint{AP-GR 66}

\date{\today}
\maketitle
\section{Introduction}\label{sec1}
Recently, higher dimensional black holes have attracted much attention in the context of string theory 
and the brane world scenario.
In particular, the black ring solution~\cite{Emparan:2001wn} is one of the most important discovery
because that means the uniqueness theorem does not hold in higher-dimensional space-time 
unlike the case of four-dimensional space-time\footnote{
However, some partial achievements was obtained in~\cite{Morisawa:2004tc,Hollands:2007aj,Hollands:2007qf,Morisawa:2007di}.}
and 
the shape of black objects can take various topology in higher-dimension.
In fact, many solutions which have more complicated structure have been
constructed~\cite{{Pomeransky:2006bd},Elvang:2007rd,Iguchi:2007is,Evslin:2007fv,Izumi:2007qx,Elvang:2007hs,Elvang:2004rt,Gauntlett:2004qy,Ishihara:2005dp,Gauntlett:2002nw,Ishihara:2006iv,Ishihara:2006pb,Yazadjiev:2007cd,Yazadjiev:2008pt} (see also~\cite{Emparan:2006mm,Emparan:2008eg}).

It is interesting to have black ring solution with a cosmological constant in the context of 
AdS/CFT correspondence.
However, by now, attempts to obtain regular black ring solution 
with a cosmological constant did not succeed~\cite{Chu:2006pf,Kunduri:2006uh}.
This might be because 
the co-existence of the scales of the diameter of black ring and the cosmological constant is difficult.
In~\cite{Caldarelli:2008pz}, Caldarelli et al. constructed solutions for thin black rings 
in dS and AdS space-times using approximate methods,
and they mentioned static black ring can exist in the case of positive cosmological constant 
because of the force balance between gravitational force and repulsive force by cosmological constant.

We consider a possibility that  
solution is dynamical by the existence of positive cosmological constant unlike the other works so far.
In general relativity, it is difficult to obtain dynamical black hole solutions,
however we can easily construct such the black hole solution in the case of the mass equal to the charge
as is constructed by Kastor and Traschen~\cite{Kastor:1992nn}.\footnote{
If we take the cosmological constant to zero, the solution~\cite{Kastor:1992nn}
reduces to the Majumdar-Papapetrou solution~\cite{Majumdar:1947eu,Papaetrou:1947ib,Hartle:1972ya} 
which describes static multi-black holes.
The construction of the solution is
possible because of a force balance between the gravitational and Coulomb forces.
The higher-dimensional generalizations of the multi-black holes are discussed in
\cite{Myers:1986rx,Gauntlett:2002nw,Gauntlett:2004qy,Ishihara:2006iv,Ishihara:2006pb}, and recently
the smoothness of horizons of higher-dimensional multi-black holes are also investigated
in \cite{Welch:1995dh,Candlish:2007fh,Candlish:2009vy,Kimura:2008cq}.
}
Kastor-Traschen solution~\cite{Kastor:1992nn} was generalized to higher-dimension in~\cite{London:1995ib}
and coalescing black holes on Gibbons-Hawking space~\cite{Gibbons:1979zt} in~\cite{Ishihara:2006ig,Yoo:2007mq}.
In this paper, we discuss
the dynamical black ring solution in a variation of the Kastor-Traschen solutions.

The organization of the paper is as follows. 
The method for constructing dynamical black ring solutions are shown in \S\ref{sec2}. 
In \S\ref{sec3}, 
global structure of the solution in the case of a single black ring is discussed. 
Event horizon of coalescing multi-black rings are discussed in \S\ref{sec4}.
Summary and discussions are given in \S\ref{sec5}. 

\section{Construction of Dynamical Black Ring Solutions}\label{sec2}
We consider five dimensional Einstein-Maxwell system with a positive cosmological constant,
which is described by the action
\begin{equation}
S=\frac{1}{16\pi G_5}\int dx^5 \sqrt{-g} (R -4\Lambda-F_{\mu\nu}F^{\mu \nu} ),
\end{equation}
where $R$ is the five dimensional scalar curvature, $F_{\mu\nu}$ 
is the Maxwell field strength tensor, $\Lambda$ 
is the positive cosmological constant and $G_5$ is the five-dimensional Newton constant.
{}From this action, we write down the Einstein equation
\begin{equation}
R_{\mu\nu}-\frac{1}{2}Rg_{\mu\nu} +2g_{\mu \nu}\Lambda 
=
2\biggl(F_{\mu\lambda}{F_{\nu}}^{\lambda}-\frac{1}{4}g_{\mu\nu}F_{\alpha \beta}F^{\alpha\beta}\biggl),
\label{einsteineq}
\end{equation}
and the Maxwell equation
\begin{equation}
{F^{\mu \nu}}_{;\nu}=0.
\label{maxwelleq}
\end{equation}

In this system,
we consider the following metric and the gauge 1-form
\begin{eqnarray}
ds^2 &=& - H^{-2} dt^2 + H e^{-\lambda t} ds^2_{\rm E^4},
\label{met}
\\
A &=& \pm \frac{\sqrt{3}}{2} H^{-1} dt,
\label{gauge}
\end{eqnarray}
where $ds^2_{\rm E^4}$ is a four-dimensional Euclid space
and $\lambda = \sqrt{4\Lambda/3}$ 
and the function $H$ is
\begin{eqnarray}
H &=& 1 + \frac{1}{e^{-\lambda t}} \Psi,
\end{eqnarray}
and the function $\Psi$ is independent of time coordinate $t$.
As shown in \cite{Kastor:1992nn,London:1995ib}, 
if the function $\Psi$ is the solution of the Laplace equation on $ds^2_{\rm E^4}$
\begin{eqnarray}
\triangle_{\rm E^4} \Psi &=& 0,
\label{laplace}
\end{eqnarray}
the metric (\ref{met}) and the gauge-1-form (\ref{gauge})
become solutions of five-dimensional Einstein-Maxwell system with a positive cosmological constant.\footnote{
In~\cite{Ishihara:2006ig,Ida:2007vi} the case of Gibbons-Hawking base space is discussed.
}

In \cite{London:1995ib}, 
the function $\Psi$ was chosen as point source harmonics, and then the solution
describes coalescence of multi-black holes with the topology of ${\rm S}^3$.
In construct, in this paper,
we focus on the ring source solutions of (\ref{laplace}) given by
\begin{align}
\Psi &= \sum_i \frac{m_{i}}{\sqrt{(r_1 + a_i)^2 + r_2^2} \sqrt{(r_1 - a_i)^2 + r_2^2}}
+
\sum_i \frac{n_i}{\sqrt{r_1^2+ (r_2 + b_i)^2 } \sqrt{r_1^2 + (r_1 - b_i)^2 }},
\label{ringsource0}
\end{align}
where we use the coordinate of $ds_{\rm E^4}^2$ as
\begin{eqnarray}
ds^2_{\rm E^4} &=& dr_1^2 + r_1^2 d\phi_1^2 + dr_2^2 + r_2^2 d\phi_2^2,
\end{eqnarray}
and $\Psi$ satisfies 
\begin{align}
\triangle \Psi &= 
\sum_i \frac{m_i}{2\pi a_i}\delta(r_1 - a_i)\delta(r_2)
+
\sum_i \frac{n_i}{2\pi b_i}\delta(r_1)\delta(r_2-b_i).
\end{align} 

\section{global structure of single black ring solution}\label{sec3}
At first, we focus on a single black ring solution, namely the harmonics $\Psi$ takes the form
\begin{align}
\Psi &= \frac{m}{\sqrt{(r_1 + a)^2 + r_2^2} \sqrt{(r_1 - a)^2 + r_2^2}}.
\label{ringsource1}
\end{align}
In this case, 
the solution is dynamical because the topologies of the spatial cross section of the event horizon 
change from ${\rm S^2} \times {\rm S^1}$ at early time into ${\rm S^3}$ at late time as shown 
in the following subsection.
\subsection{Event Horizon}
\begin{figure}[!h]
\begin{center}
\includegraphics[width=0.35\linewidth,clip]{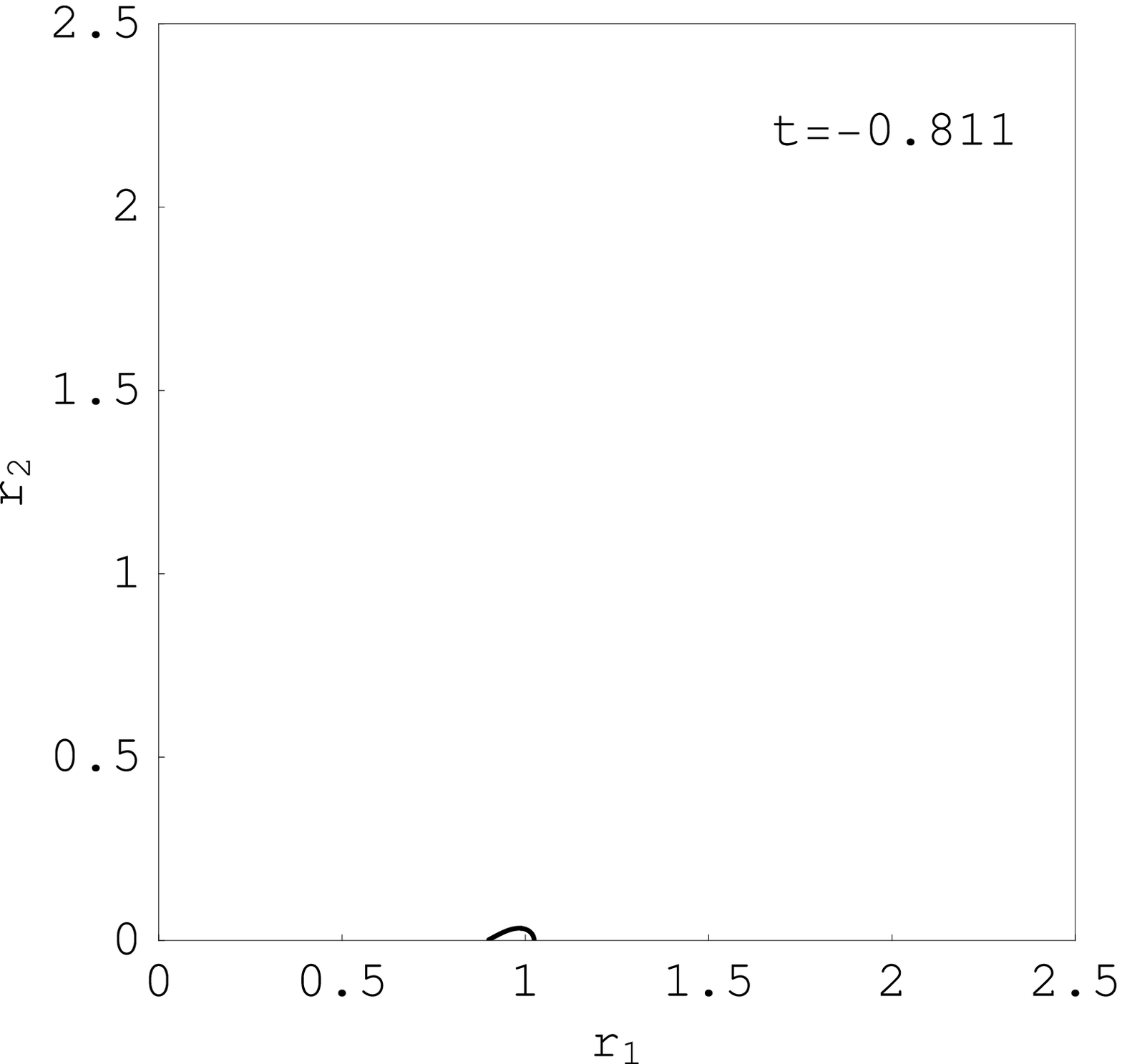}
~~~
\includegraphics[width=0.35\linewidth,clip]{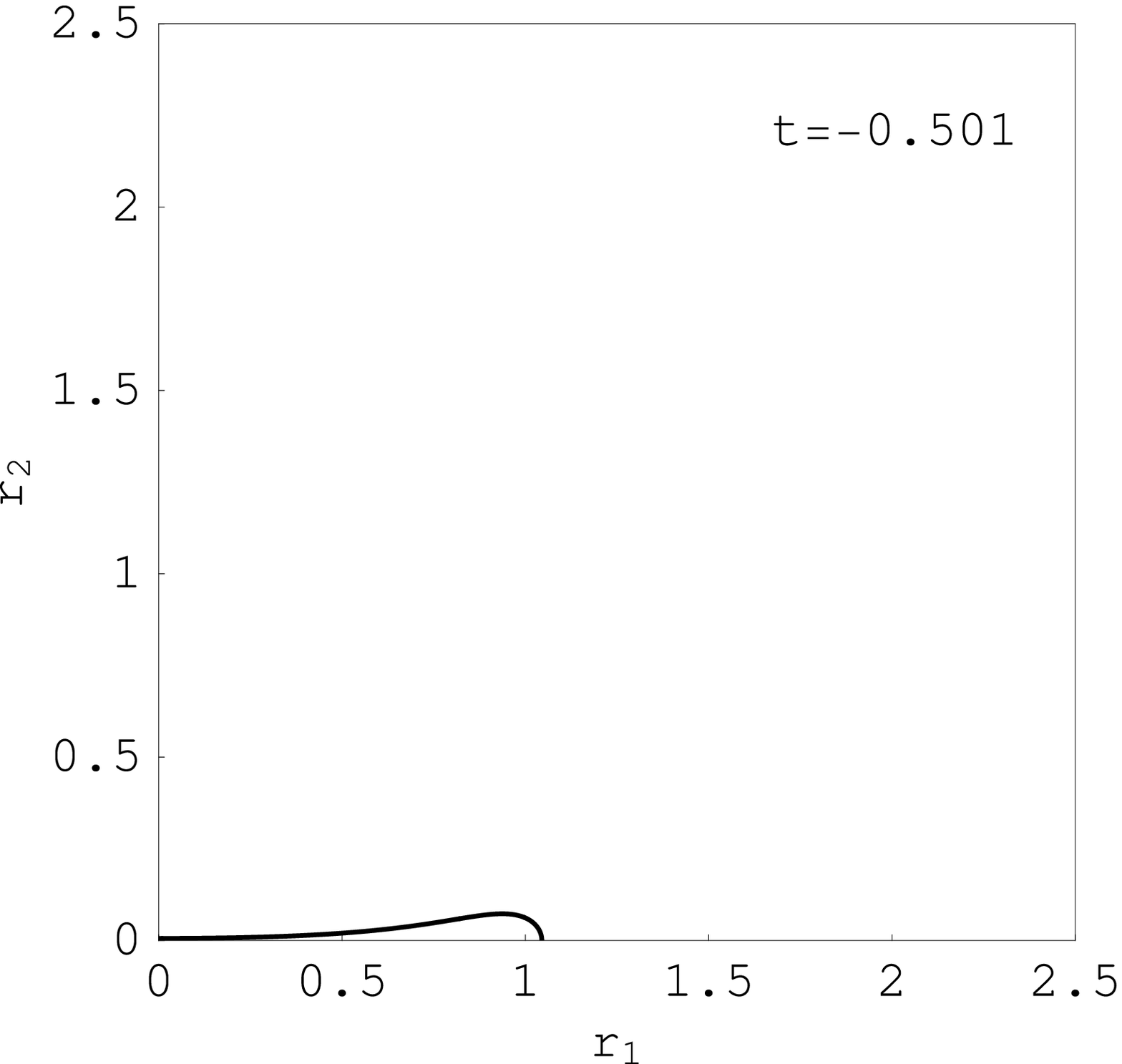}
\includegraphics[width=0.35\linewidth,clip]{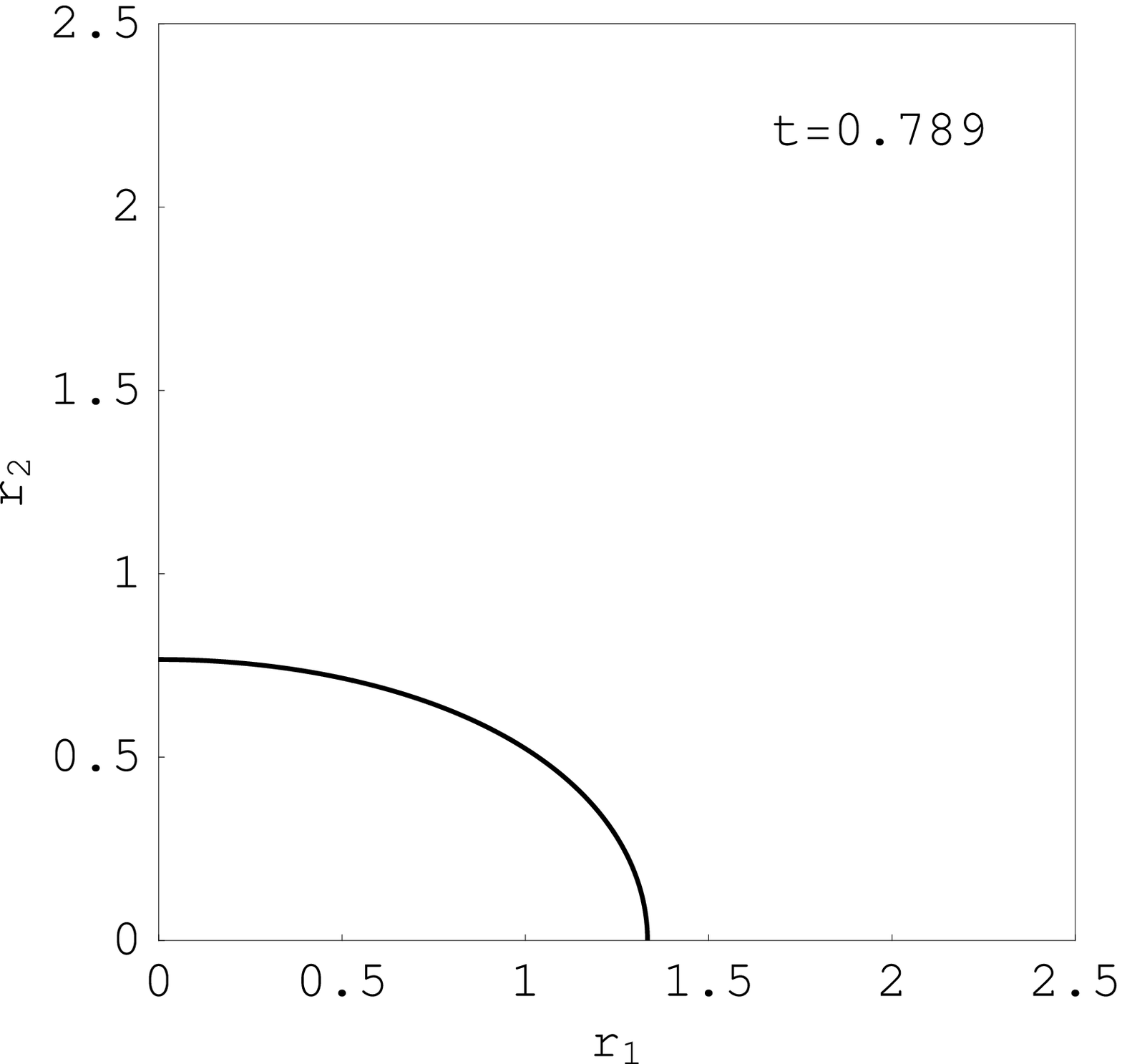}
~~~
\includegraphics[width=0.35\linewidth,clip]{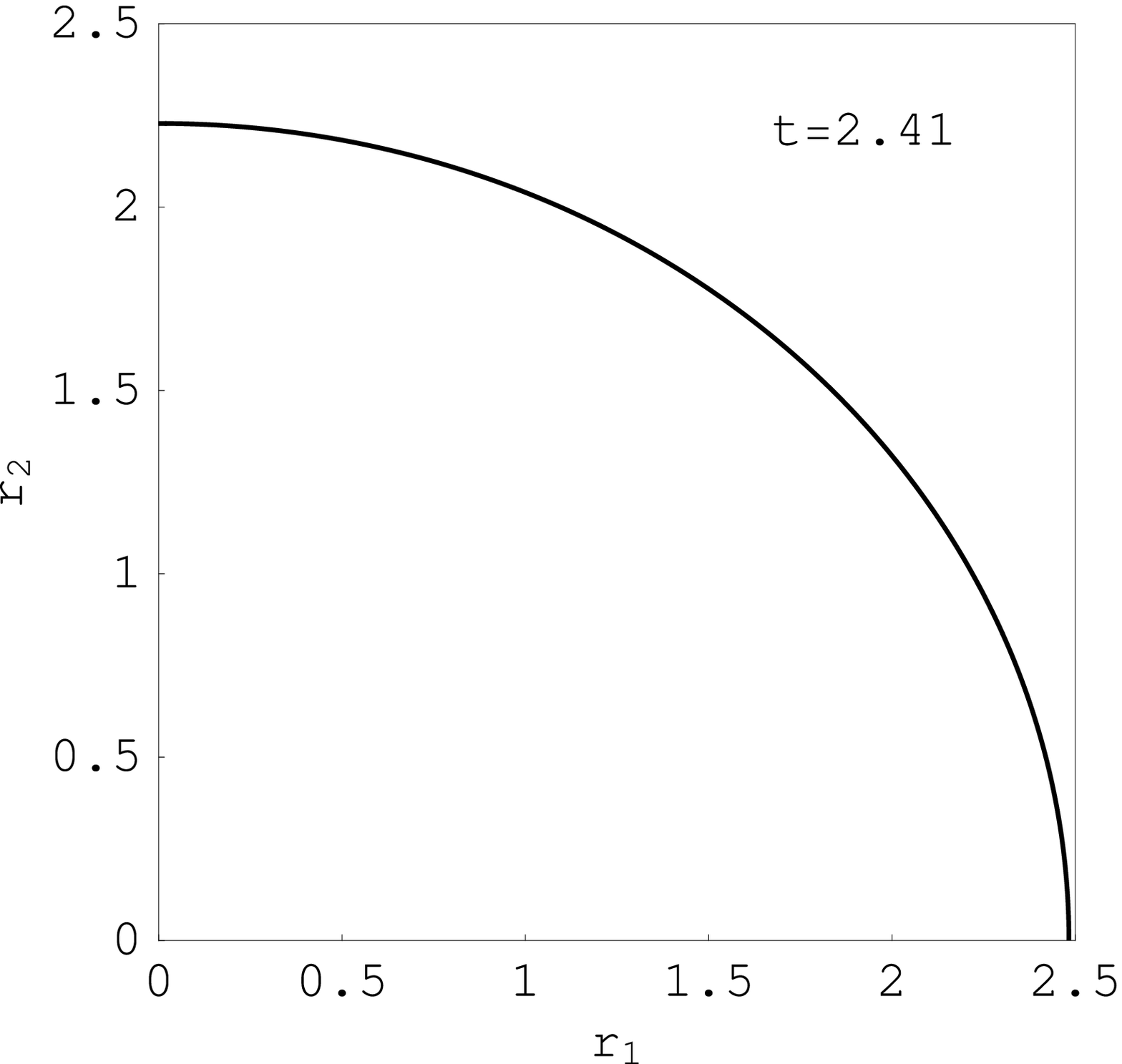}
\end{center}
\caption{
Time evolution of the event horizon for a single black ring in $r_1-r_2$ plane. 
Coordinate values of event horizon of each time slices are plotted.
We set parameters $\lambda =1,~m=1/2,~a=1$.}
\label{fig:horizon}
\end{figure}

At late time $t \to \infty$
the metric (\ref{met}) behaves 
\begin{eqnarray}
ds^2 &\simeq & -\left(1 + \frac{1}{e^{-\lambda t}}\frac{m}{r_1^2 + r_2^2}\right)^{-2} dt^2
\notag\\ & &
+
\left(1 + \frac{1}{e^{-\lambda t}}\frac{m}{r_1^2 + r_2^2}\right)e^{-\lambda t}
(dr_1^2 + r_1^2 d\phi_1^2 + dr_2^2 + r_2^2 d\phi_2^2).
\end{eqnarray}
We can see the geometry described by (\ref{met}) at late time asymptotes to that 
of Reissner-Nordstr\"om-de Sitter solution.
So we can find event horizon locally at late time if $m < m_{\rm ext} = 16/(27\lambda^2)$. 
Similar to the discussion in~\cite{Ida:1998qt},
by solving null geodesics from each point of the event horizons on $t={\rm const}.$ surface at late time
to the past, we can get null geodesic generators of the event horizons,
namely we can find the locations of the event horizons approximately.
We plot coordinate value of event horizon in $r_1-r_2$ plane at each time in Fig.\ref{fig:horizon}.
{}From this, we can see that the topology of spatial cross section of event horizon at late time is ${\rm S}^3$
and the topology of that at early time is ${\rm S}^1 \times {\rm S}^2$.

\subsection{Early Time Behavior}
{}From Fig.\ref{fig:horizon}, at early time, 
we can see the event horizon locate on near the points of ring source of harmonics
$r_1 = a, r_2 = 0$.
So in this section, we study 
the geometrical structure near $r_1 = a,~r_2=0$ analytically.
Near the points $r_1 = a, r_2 = 0$, the metric (\ref{met}) behaves 
\begin{eqnarray}
ds^2 &\simeq & -\left(1 + \frac{m}{e^{-\lambda t}} \frac{m}{2a \rho}\right)^{-2}dt^2 
\notag\\ &&
+ \left(1 + \frac{m}{e^{-\lambda t}} \frac{m}{2a \rho}\right) e^{-\lambda t}\Big(
                           d\rho^2 + \rho^2(d\theta^2 + \sin^2 \theta d\phi^2) + a^2 d\phi_1^2
                             \Big),
\label{blackstring}
\end{eqnarray}
where we introduce new coordinate $\rho,~\theta,~\phi$ as
\begin{eqnarray}
\rho \sin{\theta} \cos{\phi} &=& r_1 - a,
\\
\rho \sin{\theta} \sin{\phi} &=& r_2 \sin \phi_2,
\\
\rho \cos{\theta} &=& r_2 \cos \phi_2.
\end{eqnarray}
{}From this, we can see that the early time behavior is like black string.
The metric (\ref{met}) describes the physical process such that 
a thin black ring at early time 
shrinks  and changes into a single black hole as time increase.

If we take $\lambda = 0$ limit of the metric (\ref{blackstring}), it reduce to 
the charged black string solution with the mass equal to the charge \cite{Horowitz:2002ym}.
The charged black string solution \cite{Horowitz:2002ym}
is regular solution which has two horizon if the mass is greater then the charge,
but it has naked singularity at degenerated horizon if the mass equal to the charge.
One may worry about that the metric (\ref{blackstring}) also has naked singularity.
So, we investigate whether the singularities are hidden by the horizon, i.e.,
whether the null geodesic generators of event horizon reach $r_1=a, r_2=0$ at a finite past time.

To do this we study null geodesics in the metric (\ref{blackstring}).
The geometry of (\ref{blackstring}) has $SO(3) \times U(1)$ symmetry, so 
it is sufficient to focus on $t-\rho$ part of the metic (\ref{blackstring}).
The null geodesics $\rho = \rho(t)$ which goes inward from the future to the past 
satisfies 
\begin{eqnarray}
\frac{d \rho}{dt} 
= \frac{1}{\sqrt{e^{-\lambda t}}} \left(1 + \frac{m}{e^{-\lambda t}} \frac{m}{2a \rho}\right)^{-3/2}.
\label{geodesiceq}
\end{eqnarray}
The solution of this equation (\ref{geodesiceq}) asymptotes to
\begin{eqnarray}
\rho(t) \to \frac{1}{e^{-2 \lambda t}}\frac{\lambda^2 m^3}{2 a^3},
\label{BShorizon}
\end{eqnarray}
as $t \to -\infty$.
So, we can see that the singularity is hidden by event horizon at least finite past time.
However, in the $t\to -\infty$ limit along the event horizon (\ref{BShorizon}), 
the curvature behaves
\begin{eqnarray}
R_{\alpha \beta \mu \nu}R^{\alpha \beta \mu \nu} \sim e^{-4 \lambda t} \to \infty,
\end{eqnarray}
but we consider this singularity is not so wrong as long as we focus on the 
region in which the time coordinate $t$ takes finite values.

\section{event horizon of multi black ring solution}\label{sec4}
In this section, we investigate the event horizons of coalescing multi-black rings.
For simplicity, we restrict ourselves to the solution with two black rings.
There are two typical situation, one is concentric black rings in a plane, the other is 
orthogonal black rings.
\begin{figure}[!t]
\begin{center}
\includegraphics[width=0.33\linewidth,clip]{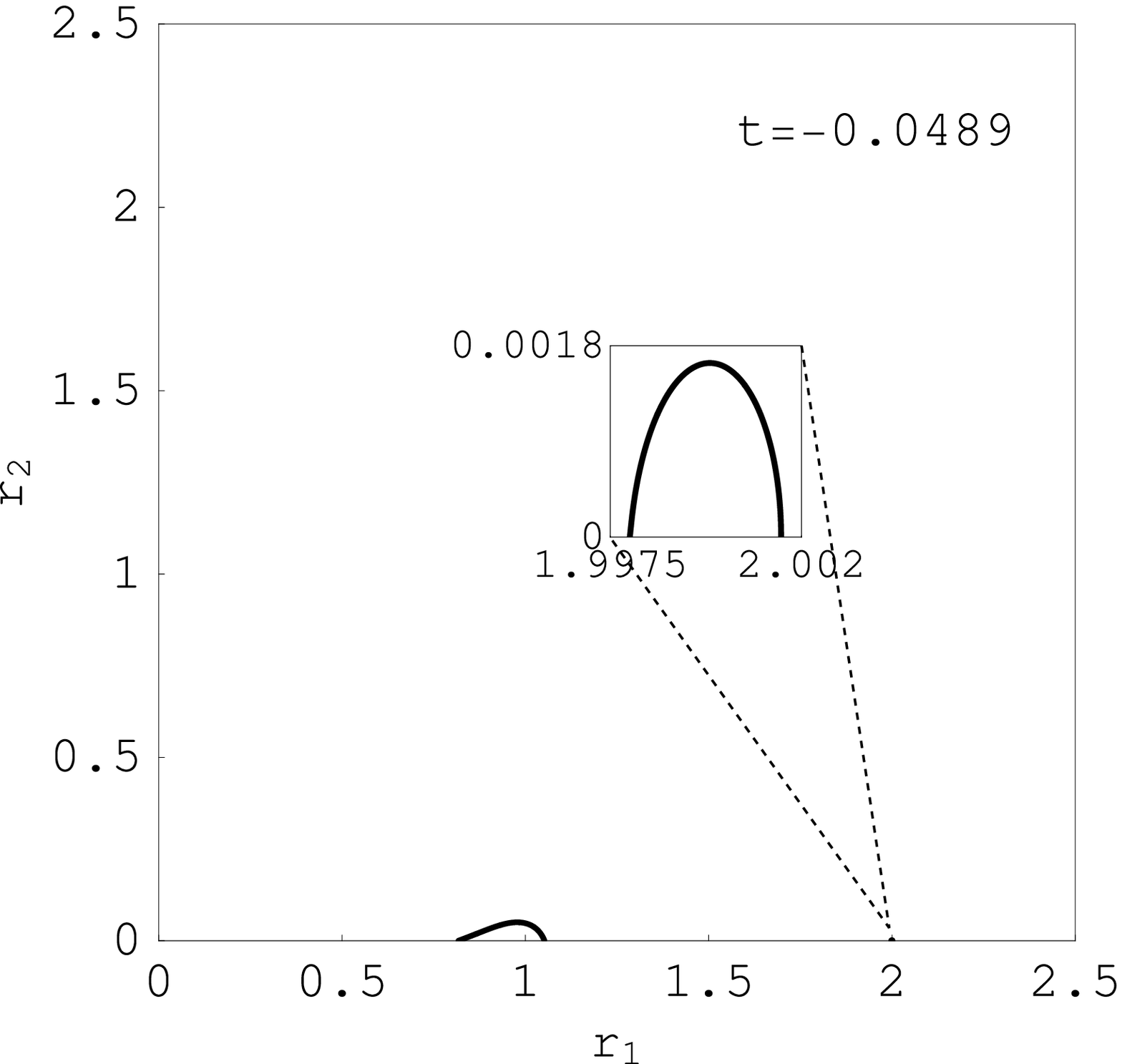}
~~~
\includegraphics[width=0.33\linewidth,clip]{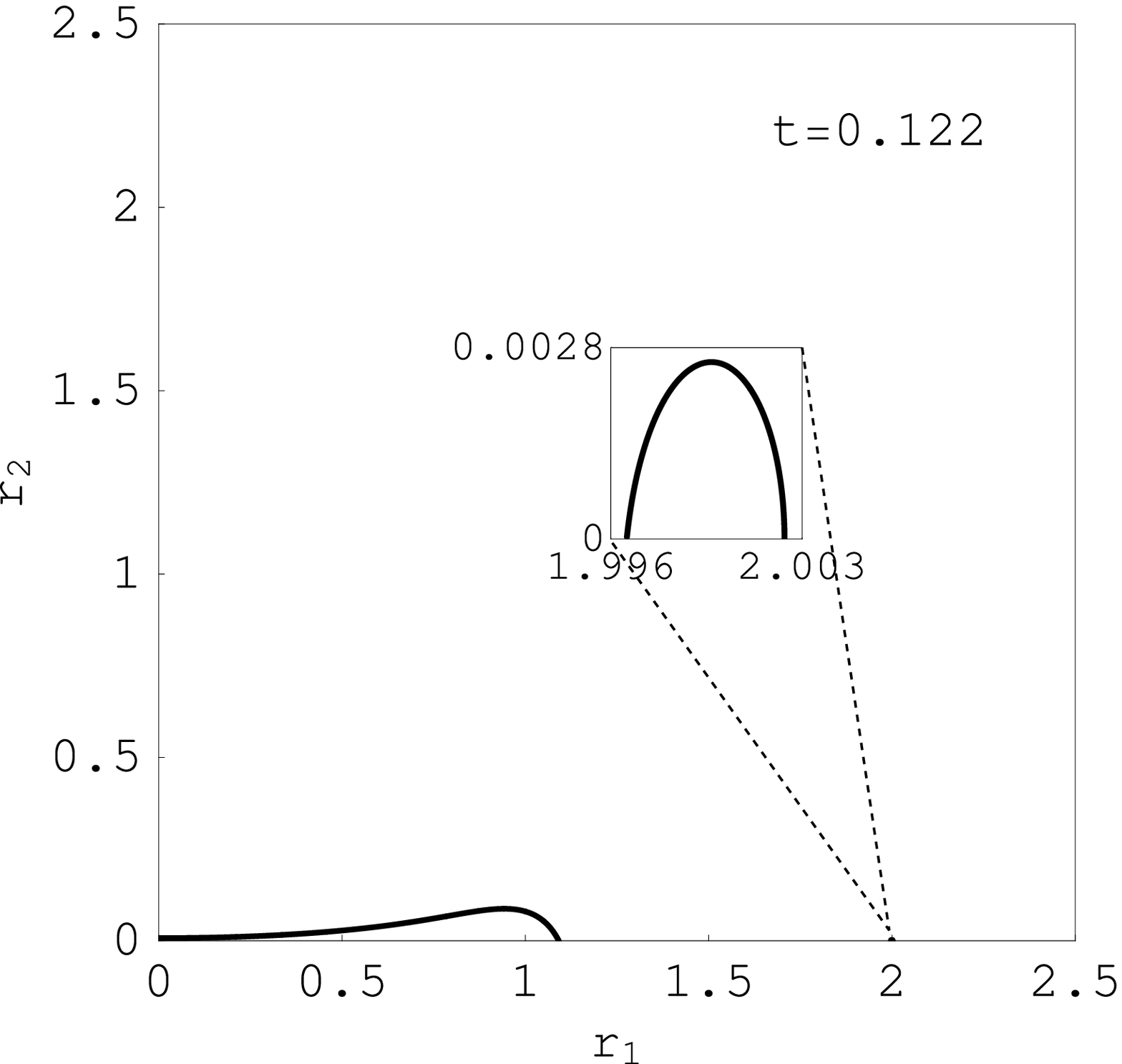}
\includegraphics[width=0.33\linewidth,clip]{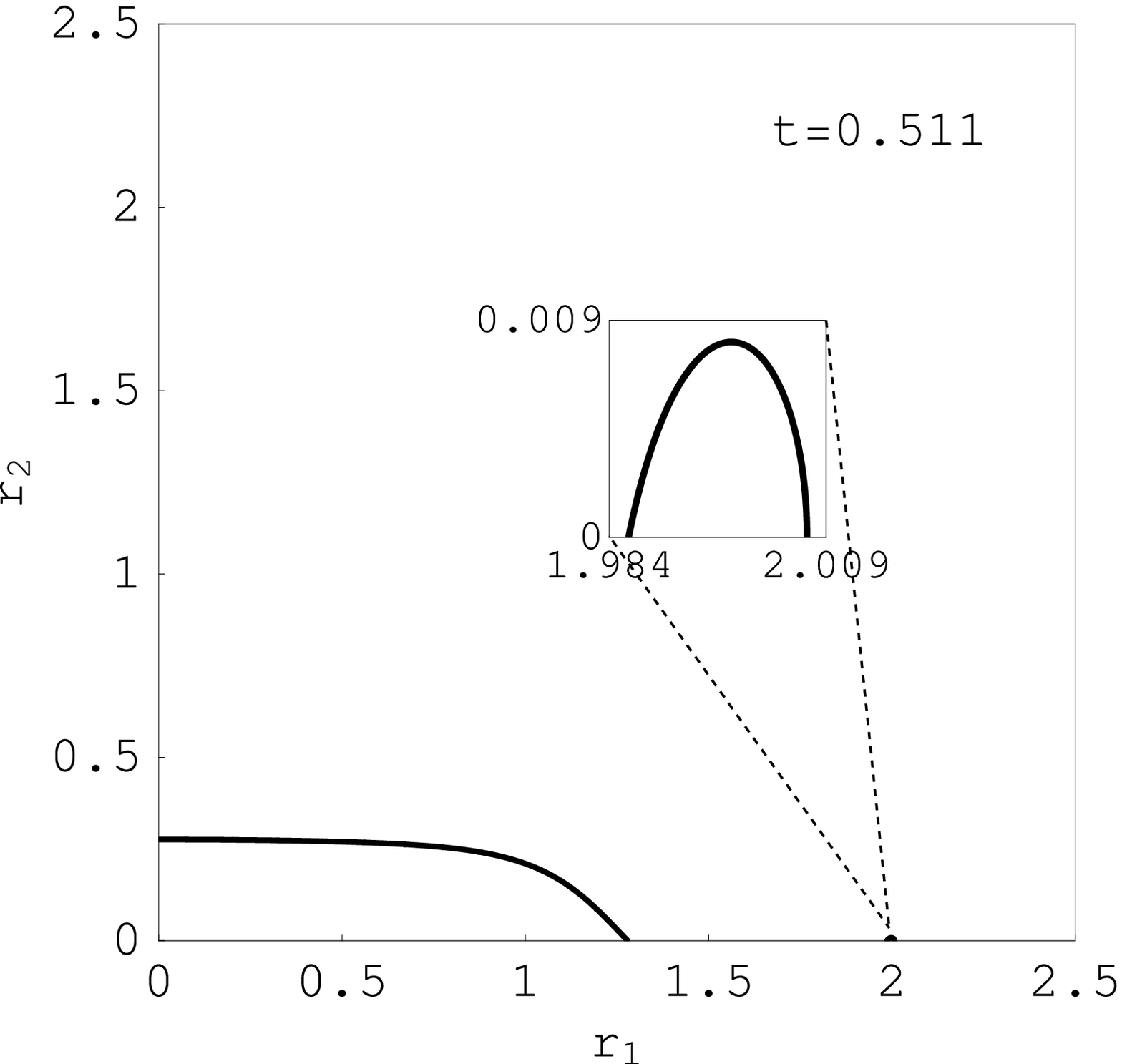}
~~~
\includegraphics[width=0.33\linewidth,clip]{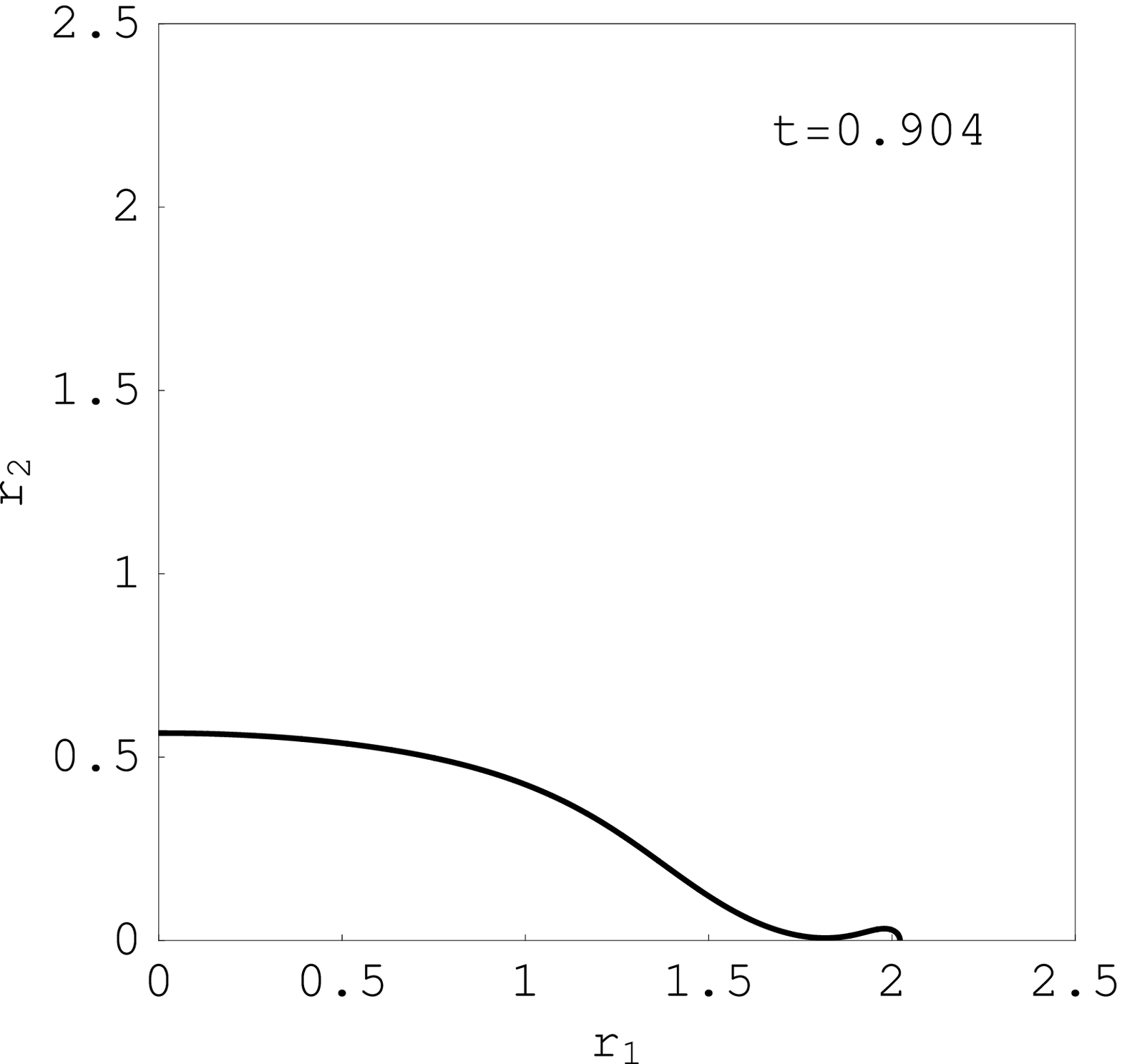}
\includegraphics[width=0.33\linewidth,clip]{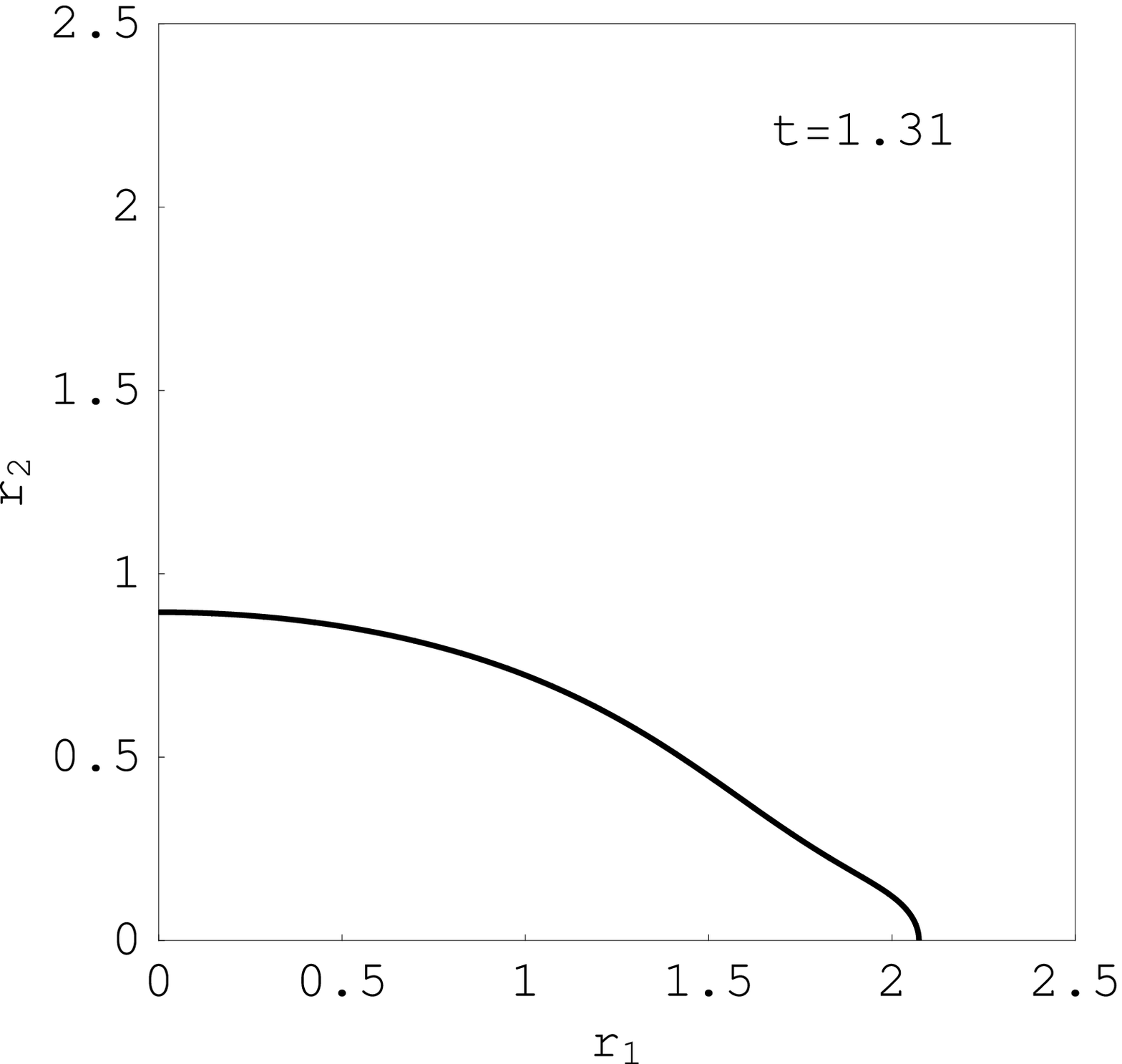}
~~~
\includegraphics[width=0.33\linewidth,clip]{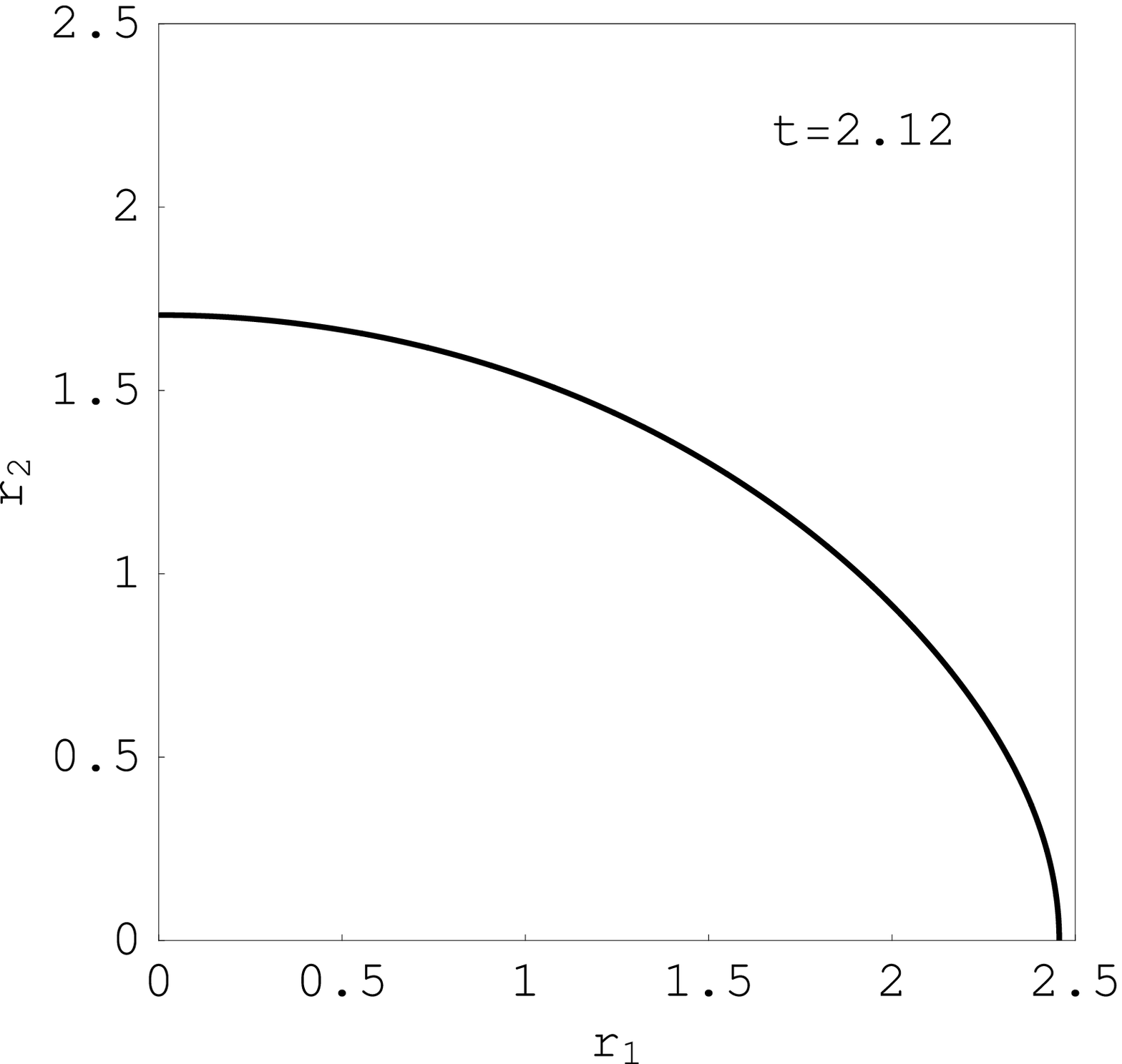}
\end{center}
\caption{
Time evolution of the event horizon for concentric black rings in a plane in $r_1-r_2$ plane. 
Coordinate values of event horizon of each time slices are plotted.
We set parameters $\lambda =1,~m_1=m_2=1/4,~a_1=1,~a_2=2$.}
\label{fig:chorizon}
\end{figure}
\begin{figure}[!h]
\begin{center}
\includegraphics[width=0.33\linewidth,clip]{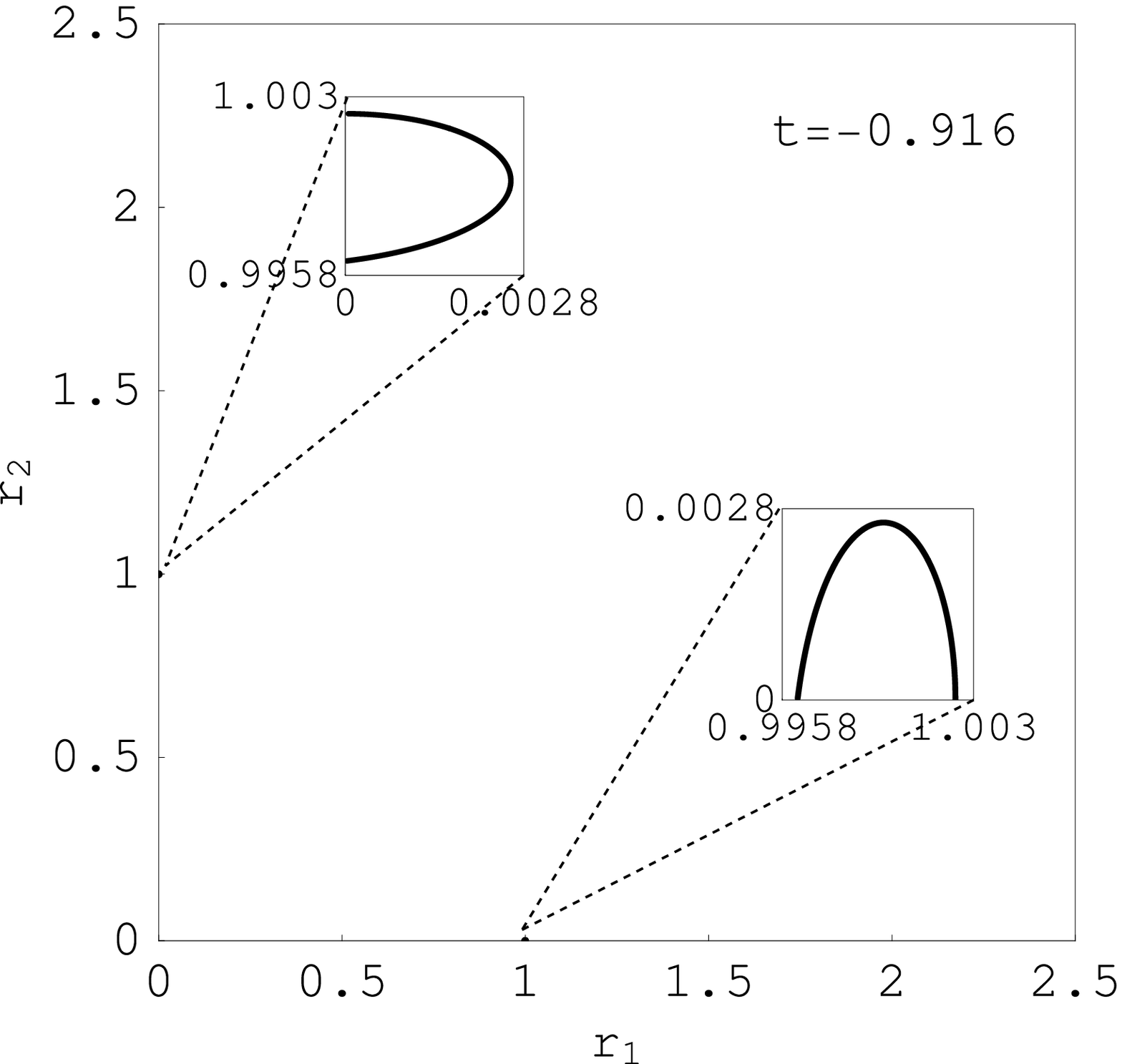}
~~~
\includegraphics[width=0.33\linewidth,clip]{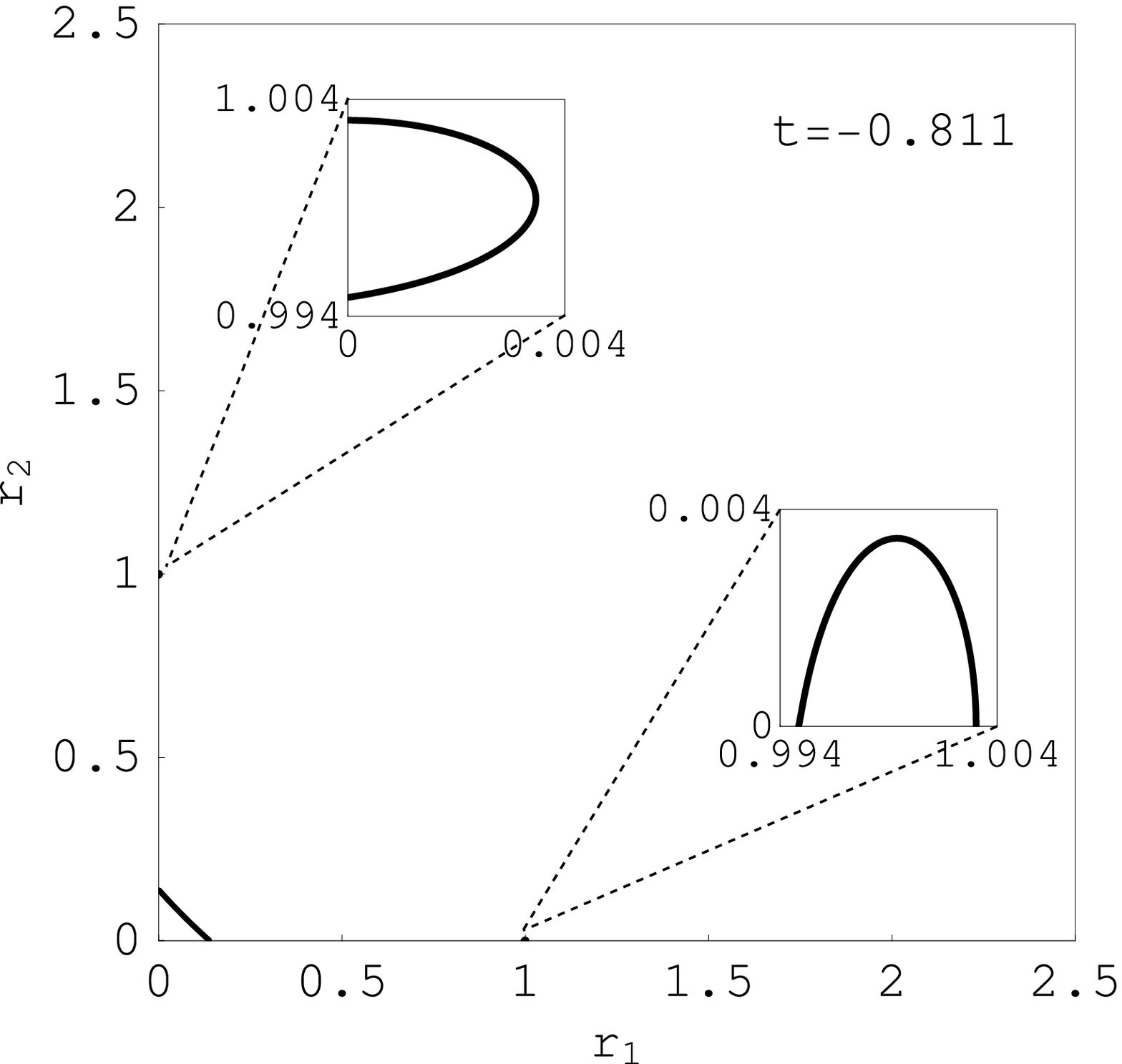}
\includegraphics[width=0.33\linewidth,clip]{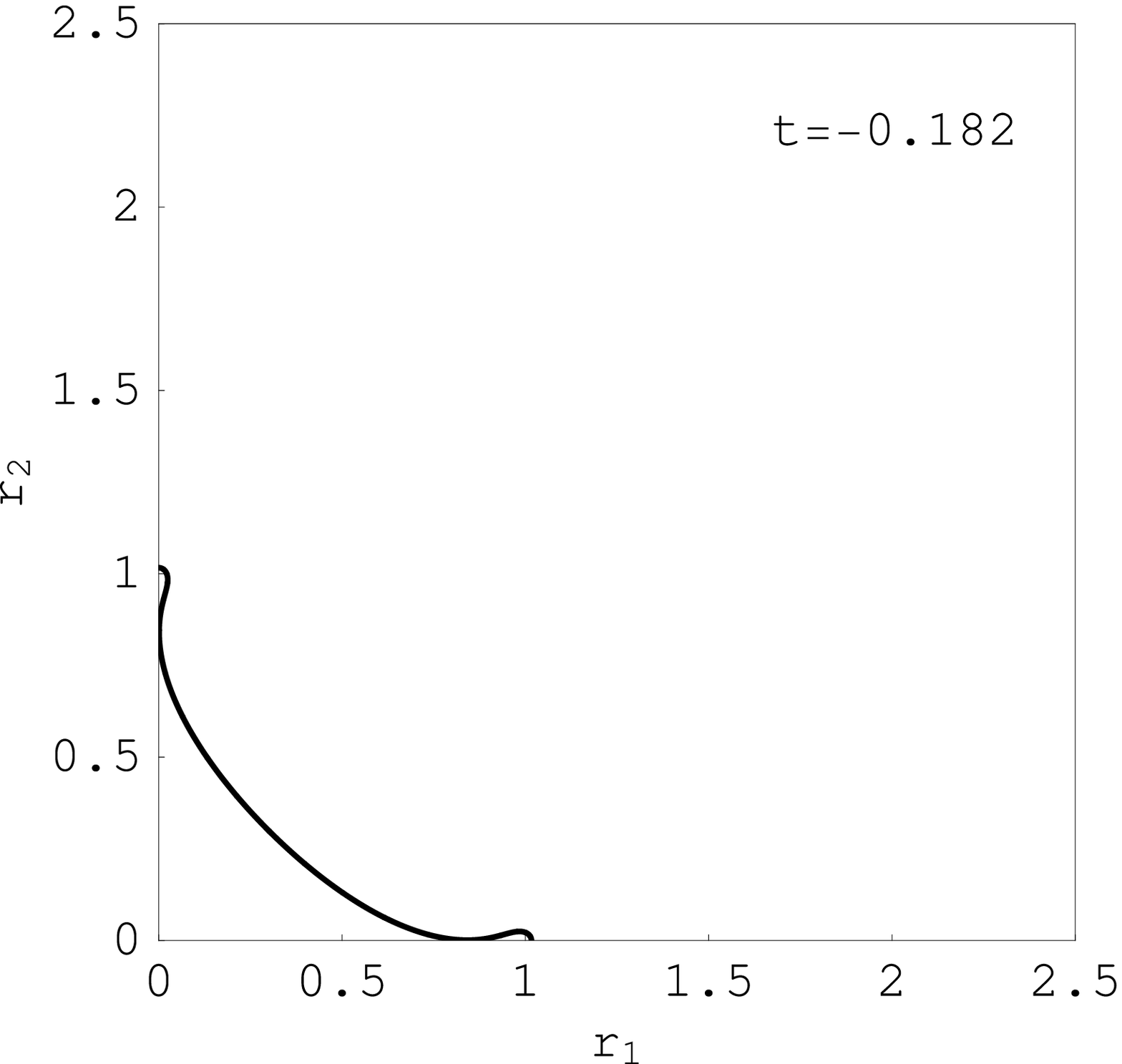}
~~~
\includegraphics[width=0.33\linewidth,clip]{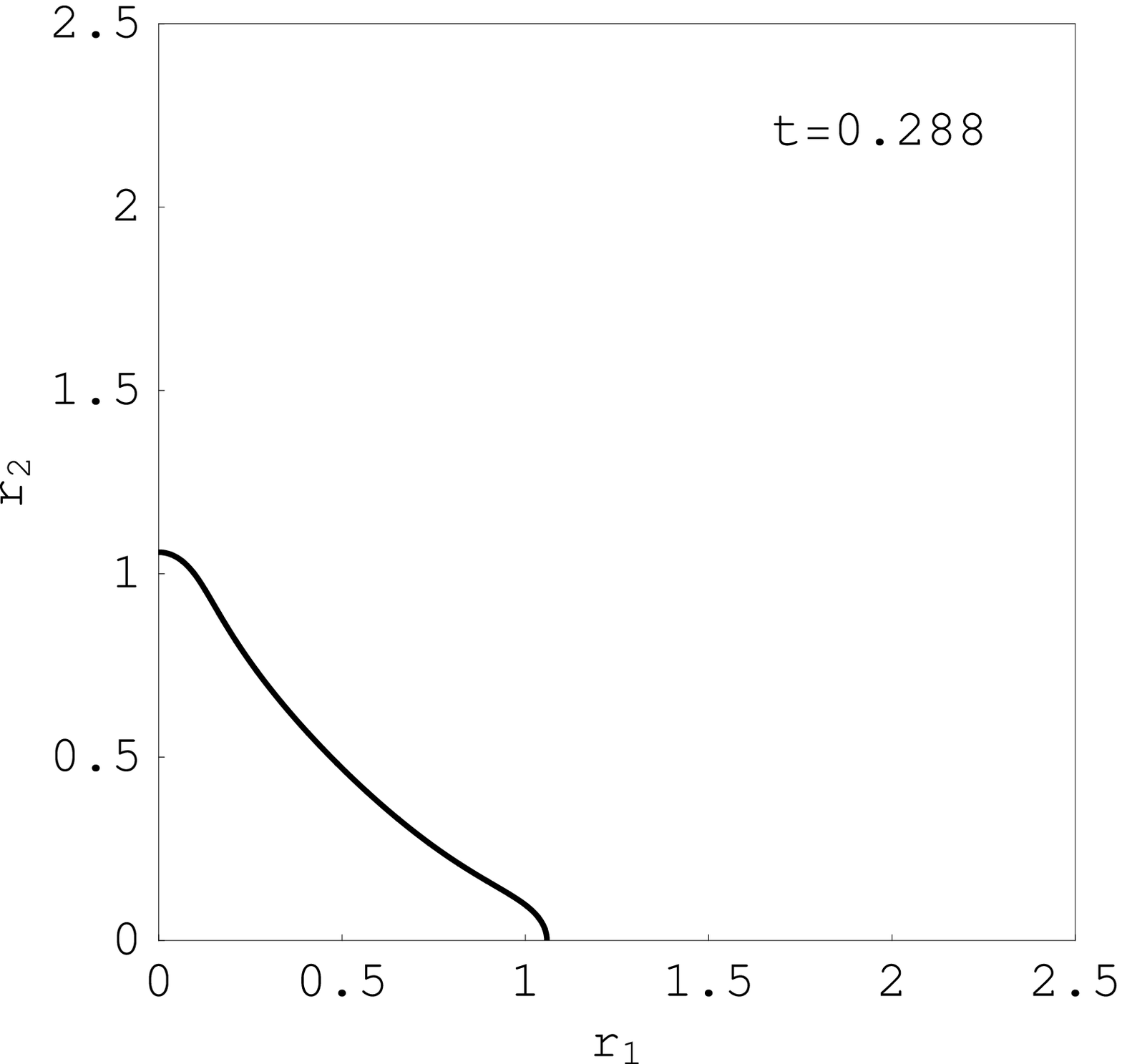}
\includegraphics[width=0.33\linewidth,clip]{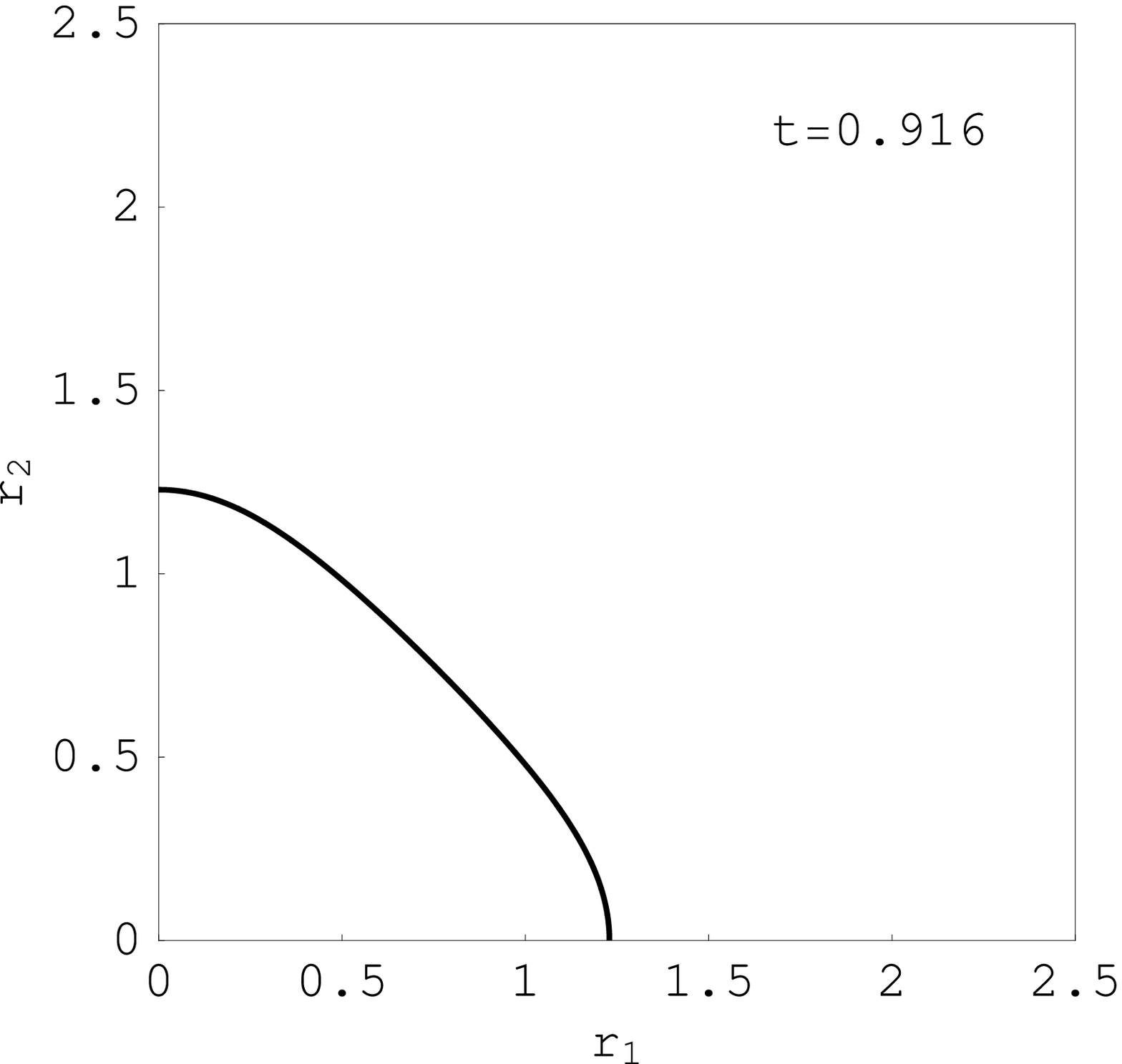}
~~~
\includegraphics[width=0.33\linewidth,clip]{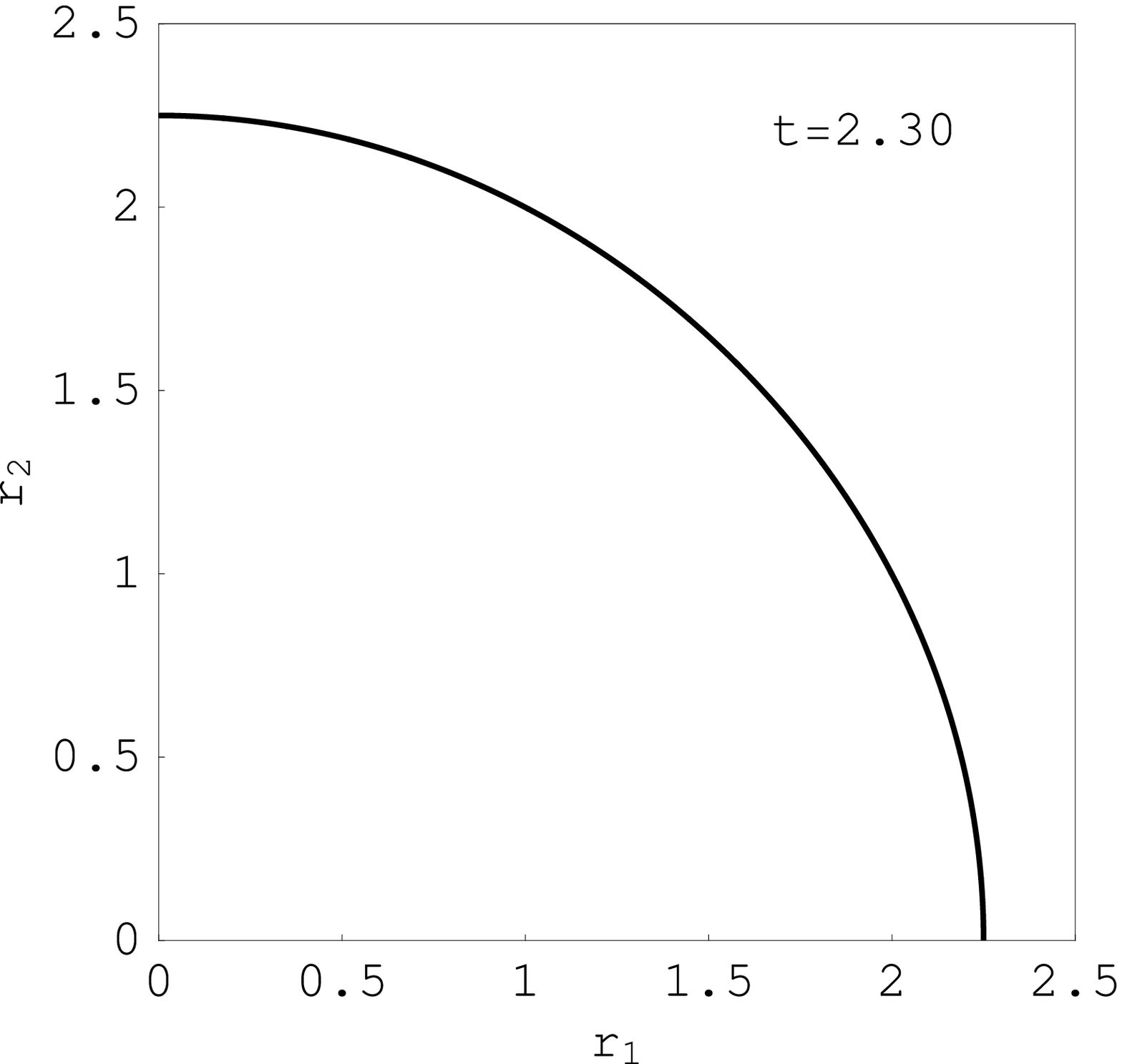}
\end{center}
\caption{
Time evolution of the event horizon for orthogonal black rings in $r_1-r_2$ plane. 
Coordinate values of event horizon of each time slices are plotted.
We set parameters $\lambda =1,~m=n=1/4,~a=b=1$.}
\label{fig:ohorizon}
\end{figure}
\subsection{concentric black rings in a plane}
In this case, $\Psi$ is given by
\begin{align}
\Psi &= 
\frac{m_1}{\sqrt{(r_1 + a_1)^2 + r_2^2} \sqrt{(r_1 - a_1)^2 + r_2^2}}
+
\frac{m_2}{\sqrt{(r_1 + a_2)^2 + r_2^2} \sqrt{(r_1 - a_2)^2 + r_2^2}},
\label{ringsource3}
\end{align}
which is constructed by two ring sources on a plane.
We can find the location of event horizon at each time slices 
plotted in Fig.\ref{fig:chorizon}
as same as the case of a single black ring solution.
At $t=-0.0489$ there are two black rings near $r_1 = 1,~r_2=0$ and $r_1=2,~r_2=0$, 
and the topology of inner black ring is changes into ${\rm S}^3$ near $t=0.122$.
The black hole with ${\rm S}^3$ and the ring near $r_1=2,~r_2=0$ coalesce into a single black hole with ${\rm S}^3$
near $t=0.904$.
\subsection{orthogonal black rings}
In this case, $\Psi$ is given by
\begin{align}
\Psi &= 
\frac{m}{\sqrt{(r_1 + a)^2 + r_2^2} \sqrt{(r_1 - a)^2 + r_2^2}}
+
\frac{n}{\sqrt{ r_1^2 + (r_2 + b)^2} \sqrt{ r_1^2 + (r_2 - b)^2}},
\label{ringsource3}
\end{align}
which is constructed by two ring sources which are orthogonal.
Similar to above discussion, 
the event horizon of this geometry is plotted in Fig.\ref{fig:ohorizon}.
At $t=-0.916$ there are two black rings near $r_1 = 1,~r_2=0$ and $r_1=0,~r_2=1$,
and at a time between $t=-0.916$ and $t=-0.811$
a black hole with ${\rm S}^3$  appears near $r_1=0,~r_2=0$.
Finally they coalesce into a single black hole with ${\rm S}^3$ near $t=-0.182$.

\section{Summary and Discussion}\label{sec5}
In this paper, we have discussed the dynamical black ring solutions
in the five dimensional Einstein-Maxwell system with a positive cosmological constant.
Our solution has constructed by use of ring source harmonics on four-dimensional Euclid space
and this is analogous to the case of 
super-symmetric black ring solution \cite{Gauntlett:2004qy}.
In the case of single ring source harmonics,
the solutions describe the physical process such that 
a thin black ring at early time 
shrinks  and changes into a single black hole as time increase.
In general, our solution can describe coalescence of multi black rings.

All regular black ring solutions so far found have angular momenta to
keep balance between gravitational force and centrifugal force, otherwise there exist some singularities.
On the other hand, our solutions do not rotate, i.e., do not have angular momentum, 
but clearly this has no conical singularity because of the way of the construction of the solution.
We consider that this is because of the balance between gravitational force and electric force.

One of important point in this paper is that 
if we set $\lambda = 0$
our solutions are static singular solutions which have curvature singularities at the points $\Psi$ diverge, 
but in the case of $\lambda \neq 0$ our solutions become regular
in the region in which the time coordinate $t$ takes finite values
 since the event horizon encloses the singularities.
This suggest that other harmonics 
which were not focused on
so far also give regular solutions with various horizon topologies.
It may be also interesting that this way applies to the case of dimensions higher than five.

While this paper was being prepared for submission,
an interesting paper \cite{Ida:2009nd} appeared, in which 
the black rings off the nuts on the Gibbons-Hawking space was discussed.

\section*{Acknowledgements}
The author would like to thank Hideki Ishihara for useful discussions.
This work is supported by the JSPS Grant-in-Aid for Scientific Research No. 20$\cdot$7858.

\end{document}